\documentclass[12pt]{article}
\usepackage{epsfig, endnotes}

\newcommand{\pdif}[2]{\ensuremath{\frac{\partial #1}{\partial #2}}}
\title{Long time simulations of granular hydrodynamics : instabilities
  and attractors} 
\author{Srikant Marakani and Gene F. Mazenko,\\
The James Franck Institute and Department of Physics,\\
University of Chicago,\\
Chicago, Illinois 60637, USA.}
\begin{document}
\maketitle
\begin{abstract}
  Using a hydrodyamic model of granular flows, we present very long
  time simulations of a granular fluid in two dimensions without
  gravity and with periodic boundary conditions in a square domain.
  Depending upon the values of the viscosity, thermal conductivity and
  dissipation, we find for intermediate times a metastable clustering
  state. For longer times the system is attracted to either a shear
  band or a vortex state.  Our results are in general agreement with
  molecular dynamics simulations.
\end{abstract}
\section{Introduction}
Granular materials are an economically important and physically
interesting system\cite{Duran}. These materials are dissipative in
nature and have complex behaviour and exhibit very interesting
phenomena such as clustering\cite{Zanetti} and {\em Maxwell Demon}
effects\cite{Eggers}.  The fact that the behaviour of these materials
is in many ways similar to conventional fluids was noted very early
and a hydrodynamic approach to them was developed heuristically by
Haff\cite{Haff} and Jenkins and Savage\cite{Savage1}. More recently a
similar hydrodynamic approach has been developed by considering a
Chapman-Enskog expansion of the Boltzmann equation for a granular
system by Brey, Dufty, Kim and Santos\cite{Dufty, Dufty2}. In this
paper, we study a modification of the Haff equations by Hill and
Mazenko\cite{ScottHill} which enabled the use of the hydrodynamic
equations to be extended to systems with clustering instabilities.
This model has the advantage of avoiding inelastic collapse and of
being able to be simulated efficiently on a computer.  Further, it was
also shown by Hill and Mazenko\cite{ScottHillone} that these
hydrodynamic equations form a well-defined system of equations under a
range of circumstances. It therefore also suggests the possibility
that the nonlinear hydrodynamical approach is applicable outside the
conventional regime of low densities.  We investigate this model in
detail and show that it exhibits many of the features found from
molecular dynamic simulations\cite{Luding, Xiaobo}. Some new features
not previously described have also been found and it should be
possible to check if these features also exist in conventional
molecular dynamic simulations.

\section{The Hydrodynamic Equations}
In this paper, we used the hydrodynamic equations postulated by
Haff\cite{Haff}. They are essentially the Navier-Stokes equations with
the transport coefficients being dependent upon the density and the
granular temperature\cite{footnote3} and with an extra dissipation
term due to the inelastic nature of the particles. Written out fully,
the equations are\cite{ScottHill}
\begin{eqnarray}
  \label{eq:haff1}
  \pdif{\rho}{t} &=& -\nabla_i (\rho u_i)\\
  \label{eq:haff2}
  \pdif{(\rho u_i)}{t} &=& -\nabla_j [(\rho T + P_f) \delta_{ij} +
  \rho u_i u_j - \eta_{ijkl} \rho T^{\frac{1}{2}} \nabla_k u_l]\\
  \label{eq:haff3}
  \pdif{T}{t} &=& -u_i \nabla_i T - T\nabla_i u_i + \frac{1}{\rho}
  \kappa \nabla_i (T^{\frac{1}{2}} \nabla_i T) + \\ \nonumber && \eta_{ijkl}
  T^{\frac{1}{2}} \nabla_i u_j \nabla_k u_l - \gamma T^{\frac{3}{2}} 
\end{eqnarray}
where sums over repeated indices is implied, $\rho$ stands for the
density, $u_i$ for the components of the velocity, $t$ for the time,
$T$ for the granular temperature\cite{footnote5}, $P_f$ for the
pressure due to the free energy, $\eta$ for the viscosity tensor,
$\kappa$ for the thermal conductivity and $\gamma$ for the thermal
dissipation coefficient. The derivation of these equations is
discussed in detail in Haff\cite{Haff}\cite{footnote2}.  We use the
usual form
\begin{equation}
  \label{eq:viscosity}
  \eta_{ijkl} = \eta(\delta_{ik} \delta_{jl} + \delta_{il}
  \delta_{jk}) + \chi \delta_{ij} \delta_{kl} 
\end{equation}
for the viscosity tensor where $\eta$ and $\chi$ stand for the shear
and bulk viscosities respectively.

Similar equations can be derived\cite{Dufty, Cubero} by performing a
Chapman-Enskog expansion of the Boltzmann equations for an inelastic
hard sphere model about the homogeneous state.
These equations are rather similar except for some scaling differences
for $\gamma$ and $\eta$ and an extra term involving the density
gradient.  The pressure term in this model is just of the ideal gas
form.  The viscosity also has a very simple form with the bulk
viscosity being equal to the shear viscosity.  

Following Hill and Mazenko\cite{ScottHill}, we simulated the Haff
equations with a pressure term which reflects the interactions between
the grains.

\section{The simulation of the hydrodynamic equations}
The Haff equations were simulated with the same basic algorithm as in
Hill and Mazenko\cite{ScottHill}. The algorithm was improved with the
use of a superior adaptive step technique which speeded up the
simulations significantly without loss of accuracy. As in Hill and
Mazenko\cite{ScottHill}, the pressure term is chosen in the model to
account for the excluded volume effects in the granular fluid.  The
pressure, is, as usual, related to the free energy density $f$ by $p_f
= \rho \pdif{f}{\rho} - f$. The interaction part of the free energy
density used for this purpose was chosen to be of the form 
\begin{equation}
f(\rho) = K(\rho - \rho_c)^{\alpha} \Theta(\rho-\rho_c)
\end{equation}
where $\Theta$ is the step function and $\rho_c$ is the packing
density.  In Hill and Mazenko\cite{ScottHill}, $\alpha = 2$. This free
energy corresponds to particles which behave roughly like stiff
springs with regard to compression. In our work, we chose $\alpha=10$
to preserve the continuity of the higher derivatives of $f(\rho)$
which enables the higher order accuracy of the fourth-order
Runge-Kutta method to be exploited.  This roughly corresponds to
particles with a very thin soft exterior and a hard interior.  In the
simulations, it is however not difficult to verify that the precise
form of this free energy density is not important as long as the
interaction is stiff enough. The large value of $\alpha$ permits a
much smaller range of densities above the packing density. The value
$K = 4 \times 10^4$ was chosen in Hill and Mazenko\cite{ScottHill}
while we chose $K = 4 \times 10^{16}$ to adjust for the larger value
of $\alpha$.

We also added a fictional term to the free energy density
\begin{equation}
f_{v} = A \Theta(\rho_v - \rho) (\rho-\rho_v)^{10}
\end{equation} 
where $\rho_v = 10^{-3}, A=10^{30}$. This contribution to the pressure
prevents areas of extremely low density from being produced since
these cause numerical instabilities.

\begin{figure}[tb]
  \centering
  \epsfig{file=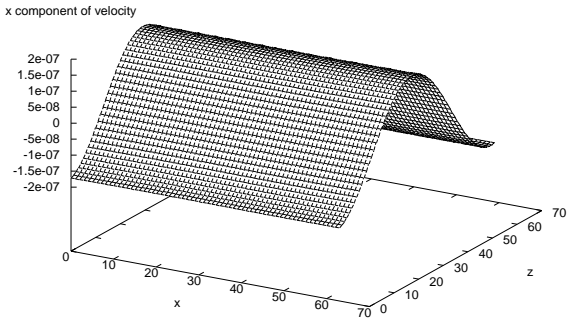, width=6cm, angle=-90}
  \caption{The $x$ component of the velocity for a system that has been
    attracted to a final shear state ($\eta=5, \gamma=1, t=10^9$).}
  \label{fig:shear}
\end{figure}

\section{Observations}
The units used in this paper are the same as in Ref. \cite{ScottHill}.
The initial conditions chosen for most of the simulations were a
uniform density half the packing density ($\rho_c = 0.2$), random
uniformly distributed initial velocities in the range $[-1, 1]$ for
each component (ensuring that the initial center of mass velocity was
zero) and a uniform granular temperature of $1$. These conditions were
chosen for their relative simplicity. The precise random configuration
of velocities was seen in some cases to matter significantly in the
subsequent dynamical evolution of the system including the form of the
final state achieved. The dynamical equations were solved numerically
on a grid in space where the lattice spacing was chosen to be 0.9 for
most simulations, a value which was verified to provide good accuracy
for the values of viscosity that we used. The densities were stored at
the intersection points of the grid while the velocities were stored
at the mid-points of the lines between the intersections
($x$-components on the lines perpendicular to the $x$-axis and the
$z$-components on the lines perpendicular to the $z$-axis). The upwind
technique was then used to determine which velocity corresponded to
which density. The method is described in detail in Ref.
\cite{Scottthesis}. The fourth order Runge-Kutta method was used for
the time evolution and an adaptive time step based on the estimated
error from a single time step and two steps of half the size used to
control the error and speed up the simulation.

There are a number of parameters in the model.  It is important to
realize that the viscosity sets the major length scale in the problem
and larger viscosities enable us to use larger lattice spacings
without introducing large numerical errors. For some low viscosity
simulations, we have used smaller lattice spacings.

\begin{figure}[bt]
  \centering
  \begin{tabular}{cc}
    \epsfig{file=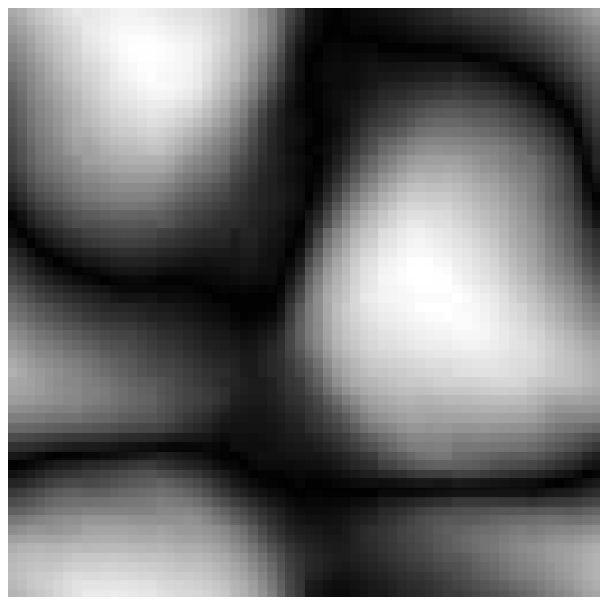, width=5cm} &
    \epsfig{file=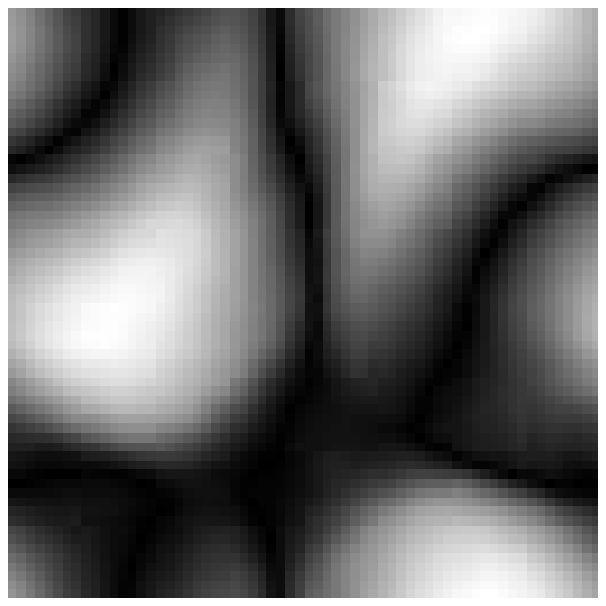, width=5cm}
  \end{tabular}
  \caption{The $x$ and $z$ velocity distributions for a vortex state seen as a greyscale map
    (intensity of white proportional to absolute velocity, $\eta=9,
    \gamma=11, t=10^9$) . For the $x$ velocity distribution shown in the
    left image, the large white region on the right has a negative
    velocity while for the $z$ velocity distribution shown in the
    right, the large white region on the left has a positive velocity.} 
  \label{fig:vortexcmap}
\end{figure}

In contrast to the work of McNamara and Young\cite{McNamara}, our
simulations have been run long enough to access the final attracting
states.  The simulations have also been performed for a wide range of
values in parameter space.  These final attracting states have been
discussed in some detail in Soto and Mareschal\cite{Soto}. There is
the well known homogenous cooling state (HCS) with uniformly
decreasing average velocity and temperature, and no structure
formation. There is the shear band configuration\cite{Soto} with the
velocity field taking on the form ${\bf v} = v_0(t) \cos(2\pi x/L)
{\bf \hat{z}}$ or ${\bf v} = v_0(t) \cos(2\pi z L) {\bf \hat{x}}$
where ${\bf \hat{z}}$ and ${\bf \hat{x}}$ are the unit vectors in the
$z$ and $x$ directions respectively and $L$ is the size of the domain.
An example of the distribution of the $x$ component of the velocity
distribution in a shear band state in the $x$ direction is shown in
figure \ref{fig:shear}.

\begin{figure}[bt]
  \centering
  \begin{tabular}{cc}
    \epsfig{file=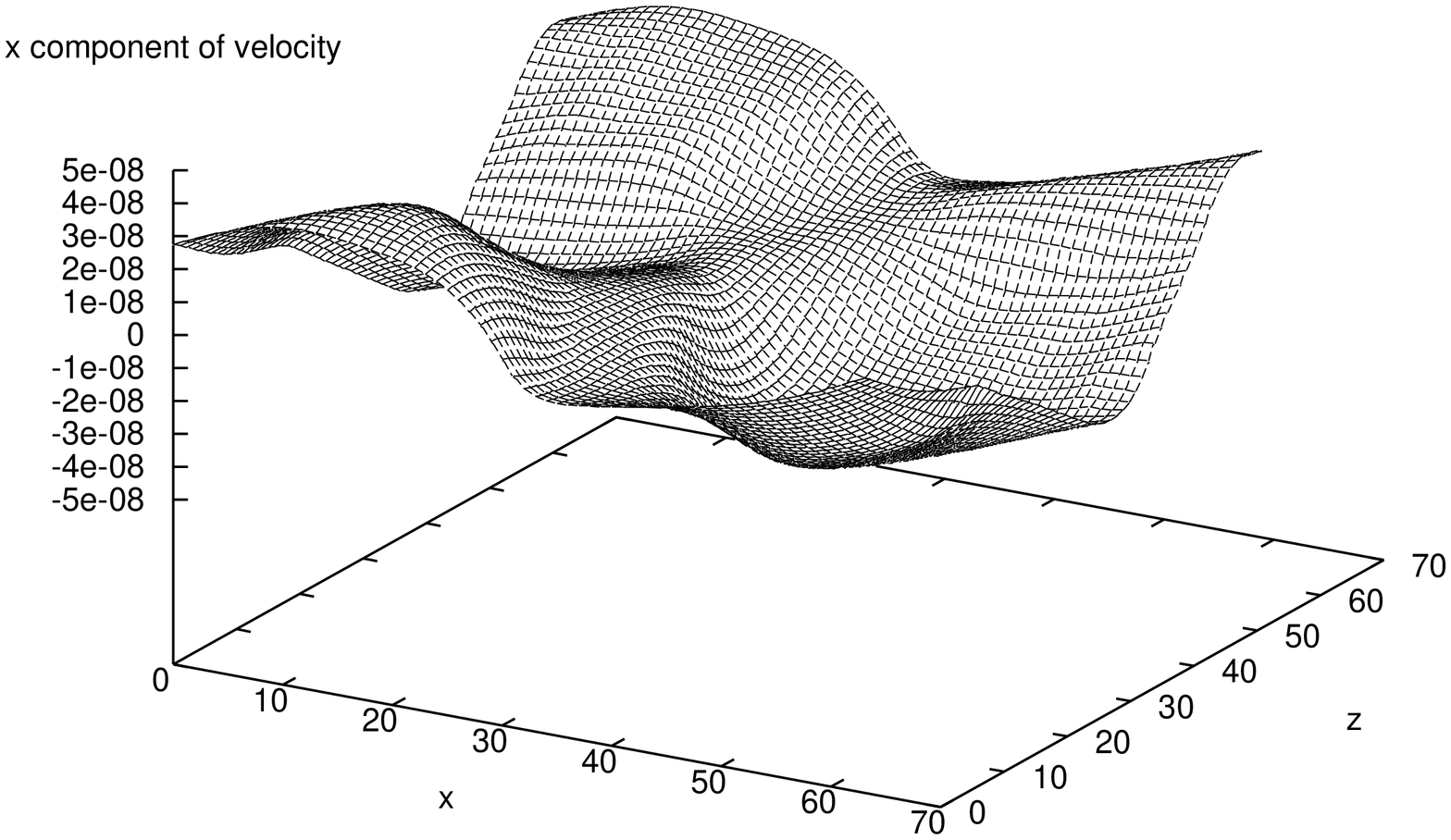, width=5cm, angle=-90} &
    \epsfig{file=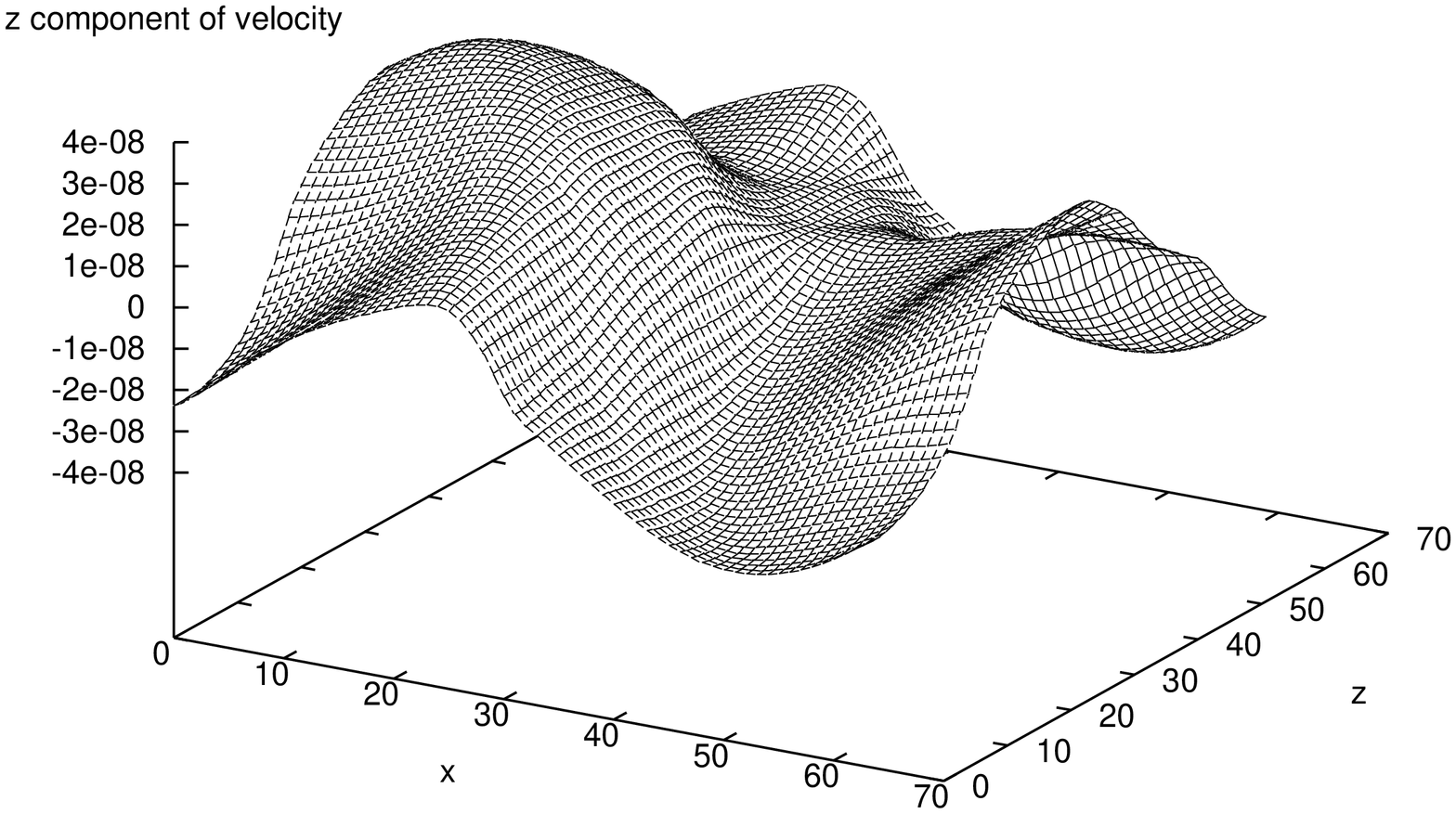, width=5cm, angle=-90}
  \end{tabular}
  \caption{The $x$ and $z$ velocity distributions for a system in the
    vortex state ($\eta=9, \gamma=11, t=10^9$).} 
  \label{fig:vortex}
\end{figure}

\begin{figure}[tb]
  \centering
  \begin{tabular}{ccc}
    \epsfig{file=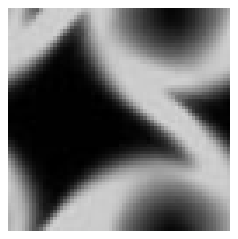, width=2cm, angle=-90} &
    \epsfig{file=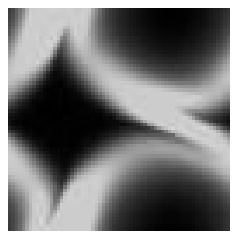, width=2cm, angle=-90} &
    \epsfig{file=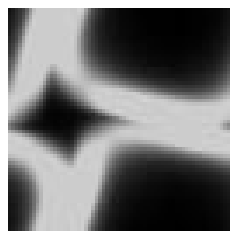, width=2cm, angle=-90}\\
    \epsfig{file=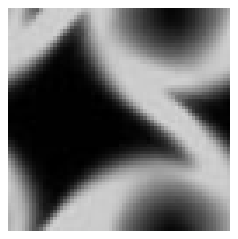, width=2cm, angle=-90} &
    \epsfig{file=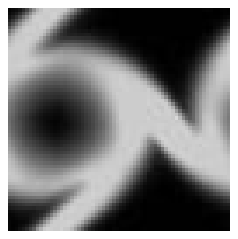, width=2cm, angle=-90} &
    \epsfig{file=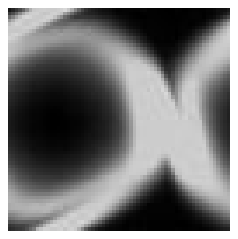, width=2cm, angle=-90}\\
    \epsfig{file=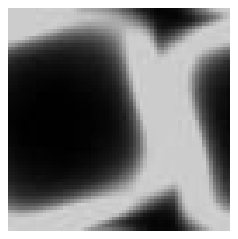, width=2cm, angle=-90} &
    \epsfig{file=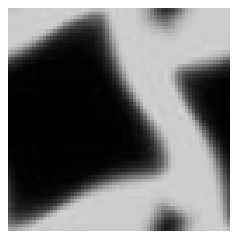, width=2cm, angle=-90} &
    \epsfig{file=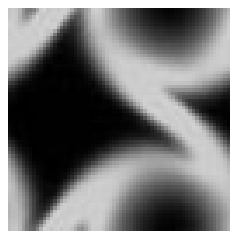, width=2cm, angle=-90}
  \end{tabular}
  \caption{The density distribution for the vortex state (density is
    proportional to the intensity of white in the figure). The time
    goes from left to right and top to bottom, with the values being
    $t=1.734 \times 10^8, 1.842 \times 10^8, 2.197 \times 10^8, 2.895
    \times 10^8, 3.779 \times 10^8, 4.545 \times 10^8, 5.437 \times
    10^8, 7.252 \times 10^8$ and $9.391 \times 10^8$.} 
  \label{fig:vortexrho}
\end{figure}

The final type of attractor is a cyclical one with features similar to
a vortex structure. If we define
\begin{equation}
  \label{eq:gamma}
  \gamma = \sqrt{\frac{E(t)}{E(0)}}
\end{equation}
where $E(t) = \langle \rho u^2 + T\rangle$ stands for the energy at
time $t$ and consider the rescaled velocities $v = u/\gamma$ and
differentially rescaled time
\begin{equation}
  \label{eq:rescale}
  ds = \gamma dt 
\end{equation}
we find that $v$ is a periodic function of $s$. A greyscale map of the
$x$ and $z$ components of the velocity distribution at a point in the
cycle is shown in figure \ref{fig:vortexcmap}.  Seen this way, the
cyclical structure looks like two tadpole-like structures moving much
like two shear bands.  This seems to be related to the state with two
counter-rotating vortices found in the nonlinear analysis of Soto and
Mareschal\cite{Soto} where the velocity is dominated by the
$k_x=\frac{2\pi}{L}, k_z = \frac{2\pi}{L}$ mode in Fourier space.
Another view of the velocity distribution is s hown in figure
\ref{fig:vortex}. The density distribution at various times in the
cycle is shown in figure \ref{fig:vortexrho}.

\begin{figure}[bt]
  \centering
  \epsfig{file=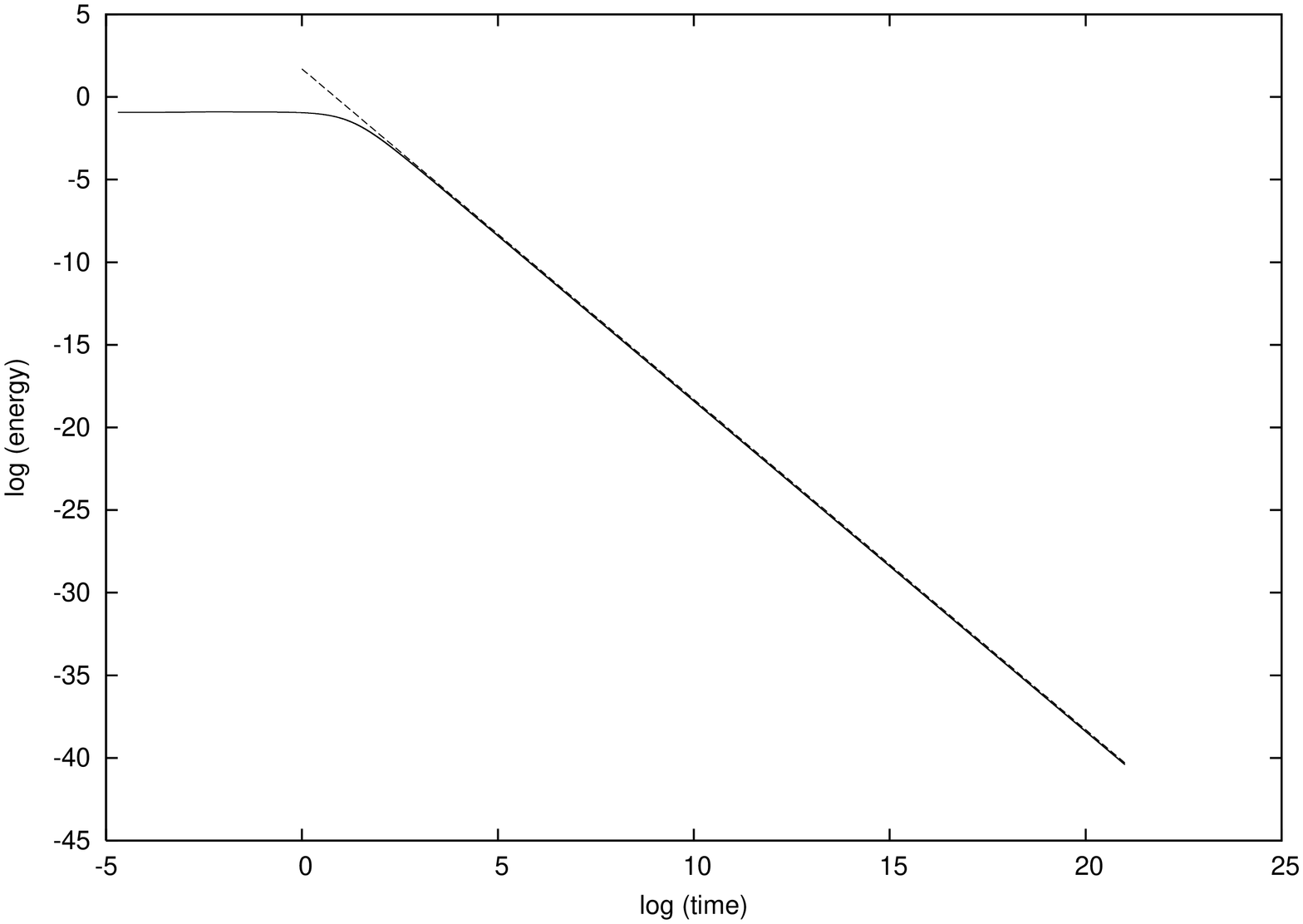, width=6cm, angle=-90}
  \caption{A plot of $\log E$ versus $\log t$ for a system that never
    leaves the HCS ($\eta=25, \gamma=0.1$). A line with slope -2 is
    drawn for reference purposes.} 
  \label{fig:hcset}
\end{figure}

\begin{figure}[tb]
  \centering
  \epsfig{file=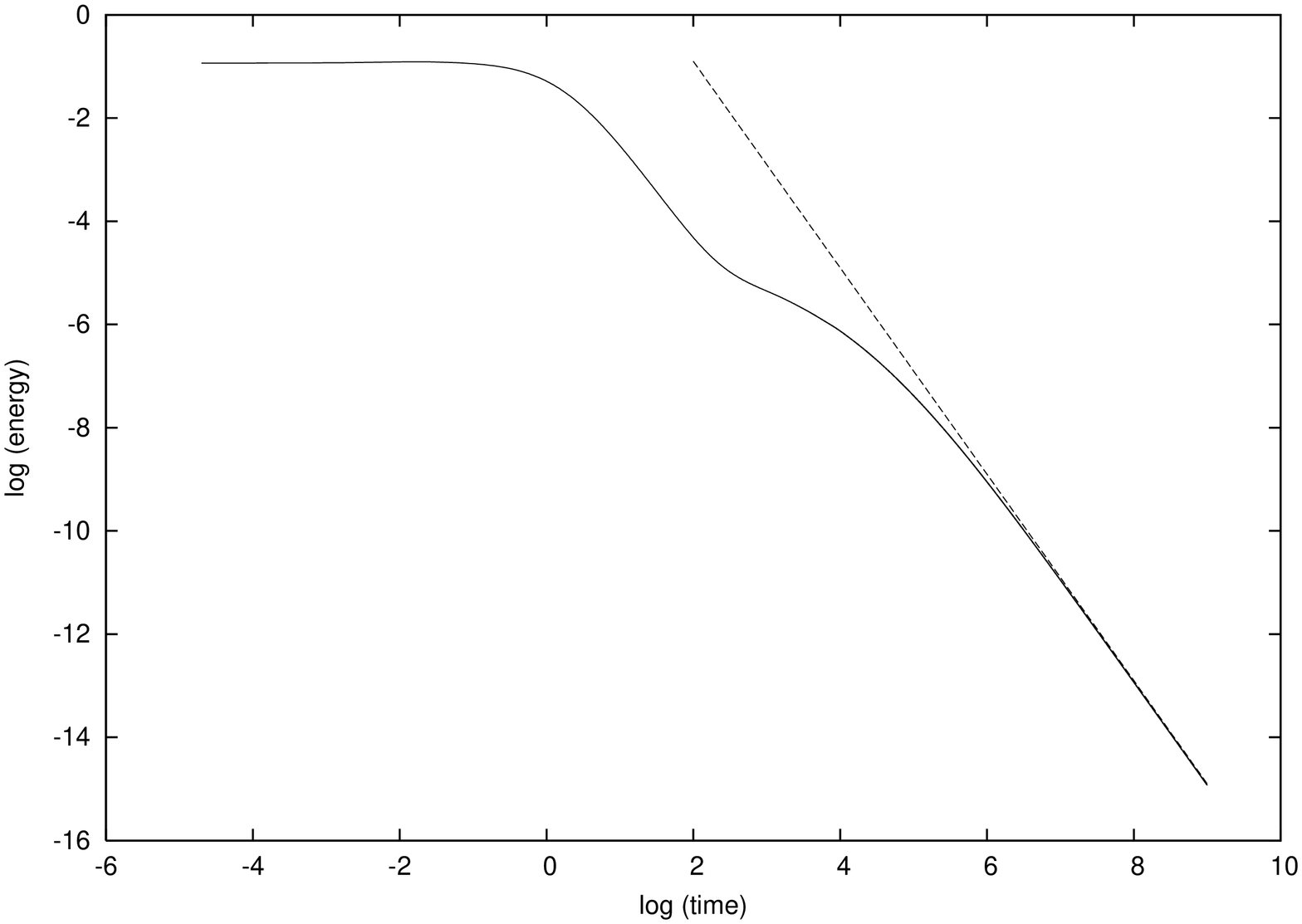, width=6cm, angle=-90}
  \caption{A plot of $\log E$ versus $\log t$ for a system that has a
    clustering instability and which finally settles into a shear band
    state ($\eta=5, \gamma=1$). A line with slope -2 is drawn for
    reference purposes.} 
  \label{fig:clusteret}
\end{figure}

We also find, as in past studies\cite{McNamara, Luding, Xiaobo}, an
intermediate clustering state which is metastable.  However, which
particular final attracting state is chosen seems to depend on many
criteria including the starting distribution of velocities.  The
homogeneous cooling state is chosen by the system when the dissipation
is low\cite{Soto}.  However, if the system size is large enough, any
dissipation will make the homogeneous cooling state
unstable\cite{Soto}. Soto and Mareschal\cite{Soto} carried out a
non-linear stability analysis for the shearing and vortex states and
found a criterion for the stability of the states provided the
deviations from the homogeneous cooling state were small.  The
criterion is relatively simple but is accurate only when the state
does not deviate too much from the homogeneous cooling state.  In our
simulations, shear bands were mostly observed but some vortex states
were also observed even with the same parameters but different system
sizes or initial conditions. For simulations with $\kappa =10,
\chi=1$, the vortex attracting state was achieved for $(\eta, \gamma)
= (7, 7), (7, 15), (9, 11)$ and $(9, 19)$ in one of the runs (since
the final state does depend on the initial conditions, it occured for
different values of the parameters in other runs).  It is also
interesting to note that the velocity field is always the first to go
to the attracting state with the other fields following much later.

\begin{figure}[h]
  \centering
  \epsfig{file=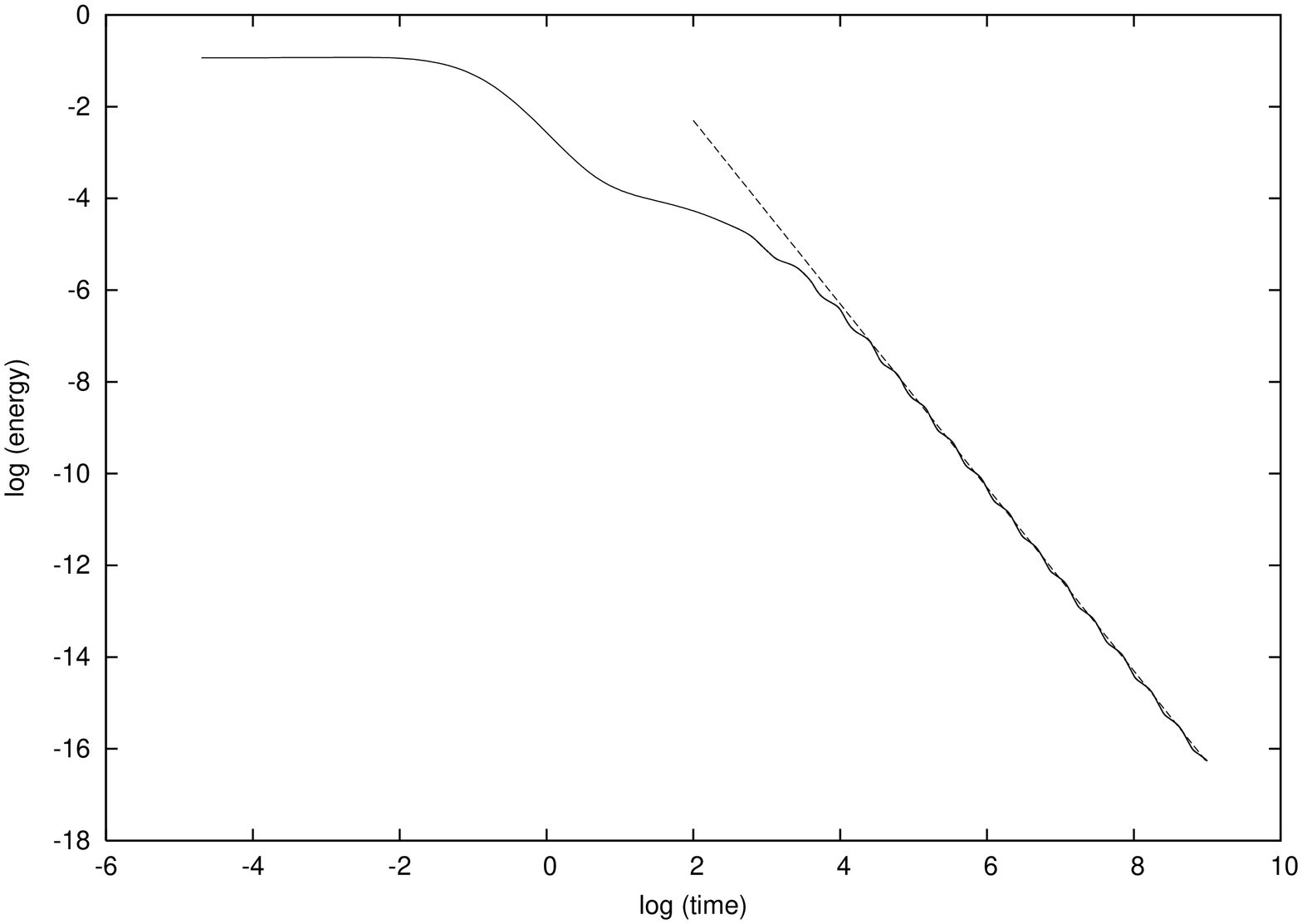, width=6cm, angle=-90}
  \caption{A plot of $\log E$ versus $\log t$ for a system that has a
    clustering instability and which finally settles into a vortex
    state ($\eta=9, \gamma=11$).} 
  \label{fig:vortexet}
\end{figure}

We can be somewhat more quantitative if we look at the evolution of
the energy (defined as $E = \langle \rho (\frac{1}{2}u^2 + T)
\rangle$) and the granular thermal energy (defined as $E_T = \langle
\rho T\rangle$) with time\cite{McNamara, Luding, Xiaobo}. It is
well known that in the HCS, the energy evolves as
\begin{equation}
  \label{eq:hcsenergy}
  E(t) = \frac{E(0)}{(1 + t/t_0)^2}
\end{equation}
so that at large times, the slope of a log-log plot of energy with
respect to time has a slope of -2. We show this plot for a system
which never leaves the HCS in figure \ref{fig:hcset}. The plot for a
system which has the clustering instability is shown in figure
\ref{fig:clusteret}. We see that during the clustering instability,
the rate of decrease of energy is reduced. Some studies\cite{Xiaobo}
indicate that the rate of decrease of the energy goes as $E \sim
t^{-1}$ during this metastable phase but we find that the relation
depends on the value of the parameters.  The plot for a system which
has the clustering instability and which settles into the vortex state
is shown in figure \ref{fig:vortexet}. The reduction in the rate of
decrease of energy in the clustering phase is similar to that of a
system that finally goes into a shear band state but the cyclical
nature of the final state is clearly seen in this plot. The average
slope of the plot in the final state is still -2.

\begin{figure}[h]
  \centering
  \epsfig{file=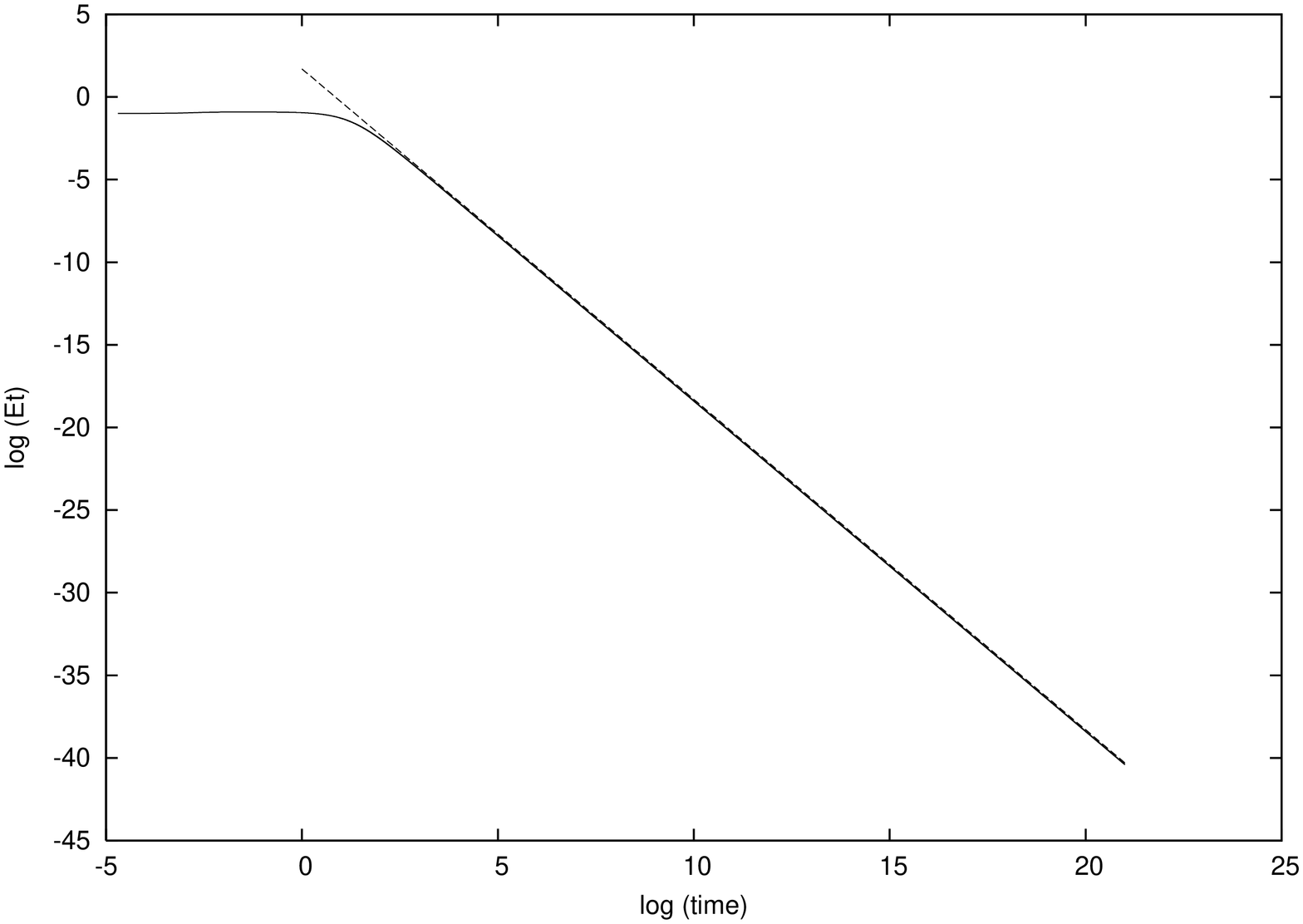, width=6cm, angle=-90}
  \caption{A plot of $\log E_T$ versus $\log t$ for a system that
    never leaves the HCS ($\eta=25, \gamma=0.1$). A line with slope 
    -2 is drawn for reference purposes.} 
  \label{fig:hcsett}
\end{figure}

\begin{figure}[h]
  \centering
  \epsfig{file=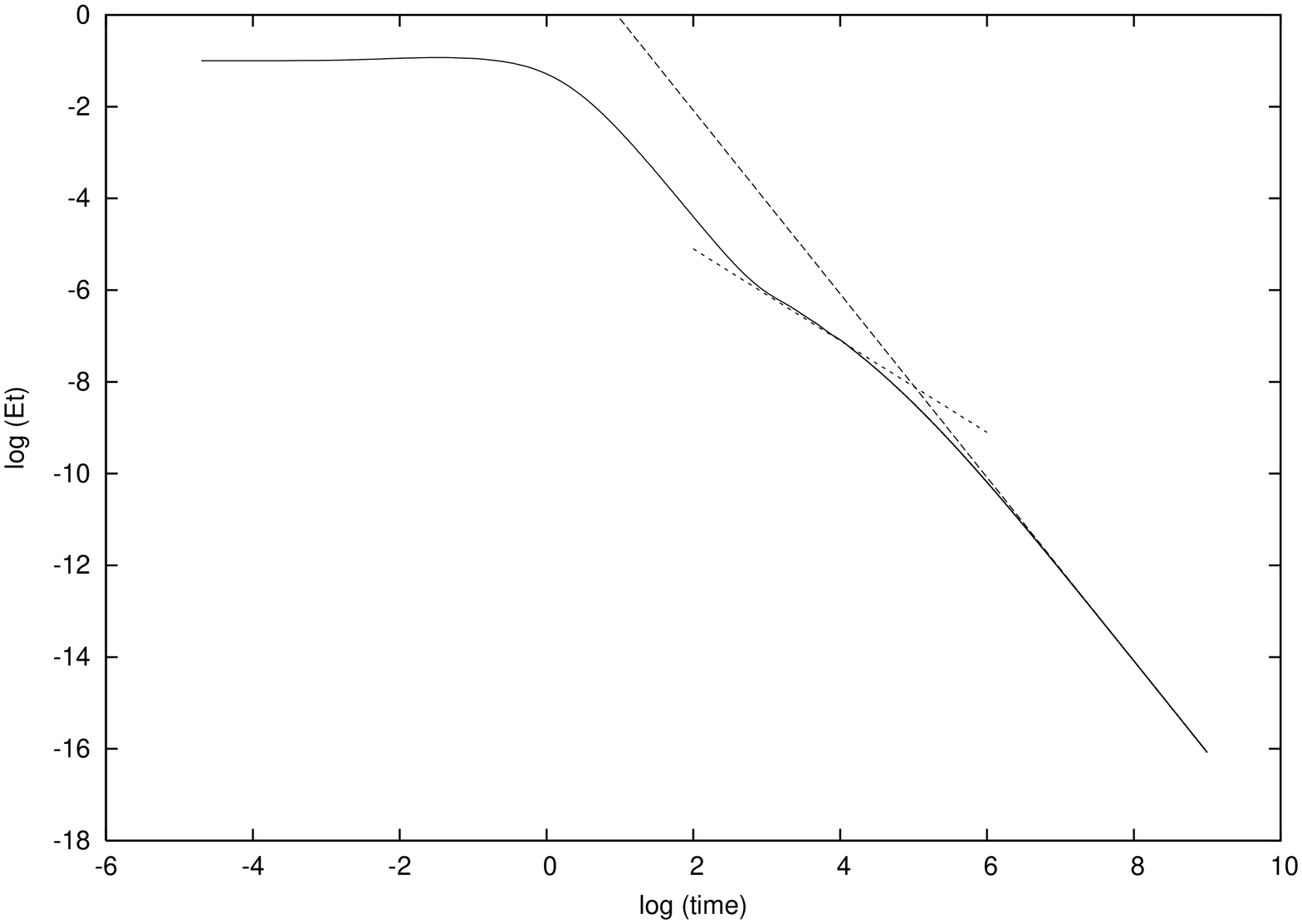, width=6cm, angle=-90}
  \caption{A plot of $\log E_T$ versus $\log t$ for a system that has a
    clustering instability and which finally settles into a shear band
    state ($\eta=5, \gamma=1$). Lines with slope -1 and -2 are drawn
    for reference purposes.}
  \label{fig:clusterett}
\end{figure}

\begin{figure}[h]
  \centering
  \epsfig{file=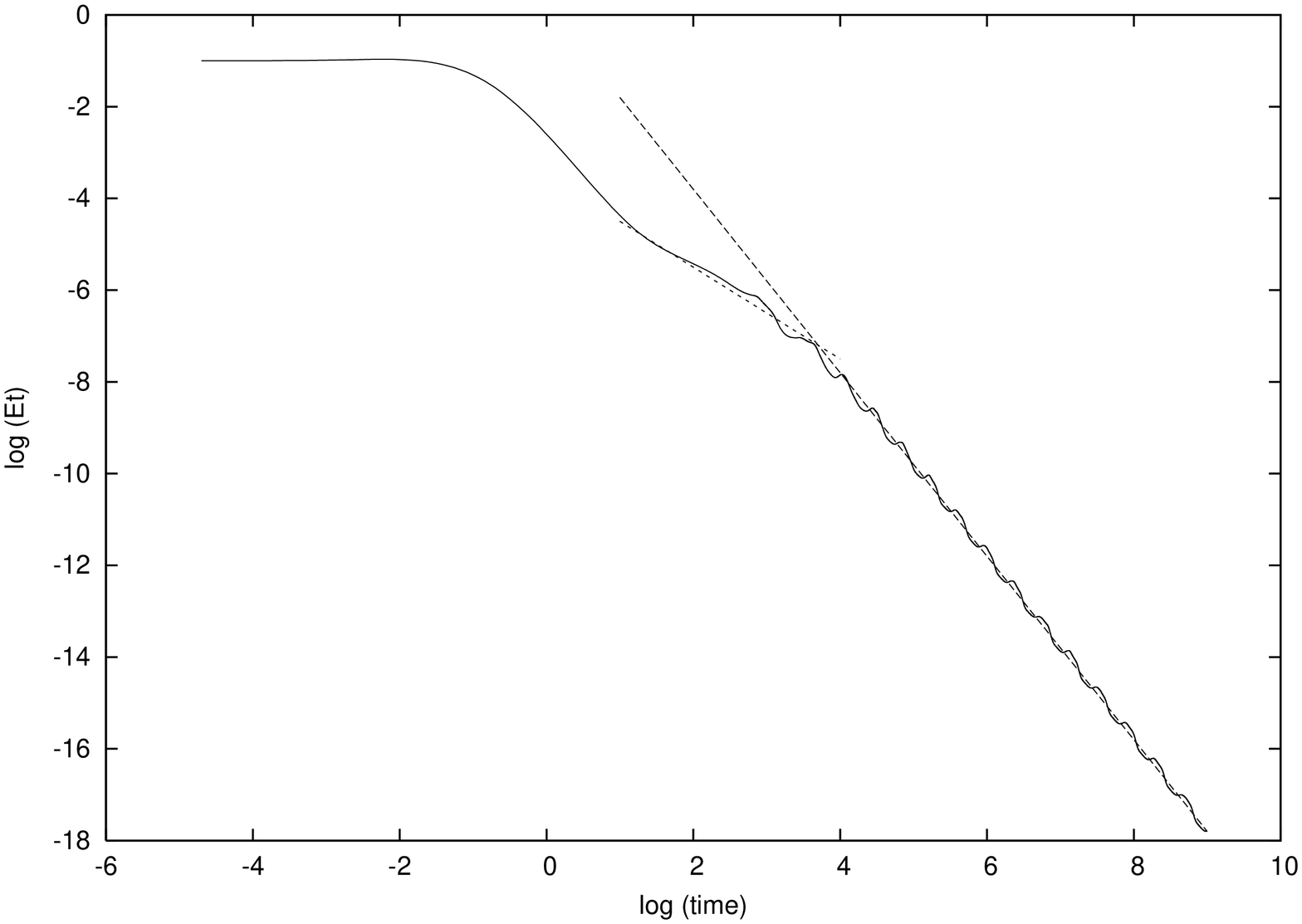, width=6cm, angle=-90}
  \caption{A plot of $\log E_T$ versus $\log t$ for a system that has a
    clustering instability and which finally settles into a vortex
    state ($\eta=9, \gamma=11$). Lines with slope -1 and -2 are drawn
    for reference purposes.}
  \label{fig:vortexett}
\end{figure}

\begin{figure}[h]
  \centering
  \epsfig{file=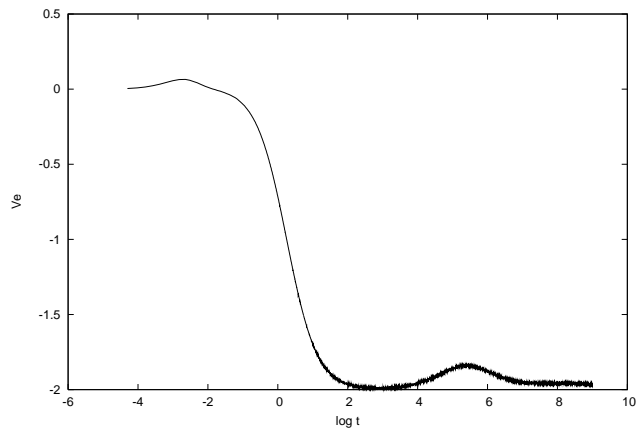, width=6cm, angle=-90}
  \caption{$V_E$ versus time for  $\eta=25, \chi=1, \gamma=1,
    \kappa=10$. This system goes directly to the shear state without a
  clustering instability.}
  \label{fig:nuE_eta25_gam1}
\end{figure}

\begin{figure}[h]
  \centering
  \epsfig{file=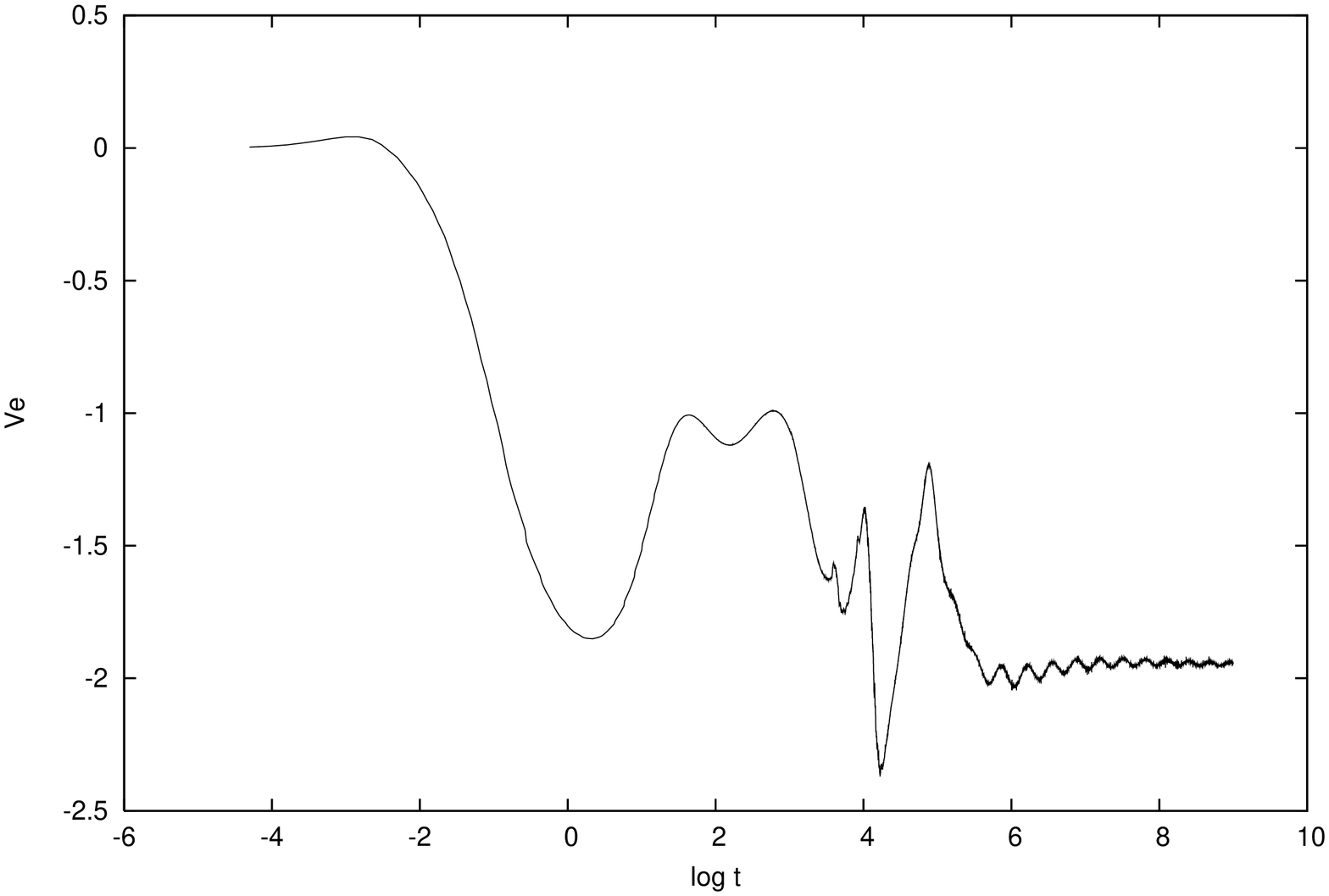, width=6cm, angle=-90}
  \caption{$V_E$ versus time for $\eta=25, \chi=1, \gamma=19,
    \kappa=10$. This state goes finally to the shear state but only
    after a clustering instability phase.}
  \label{fig:nuE_eta25_gam19}
\end{figure}

\begin{figure}[h]
  \centering
  \epsfig{file=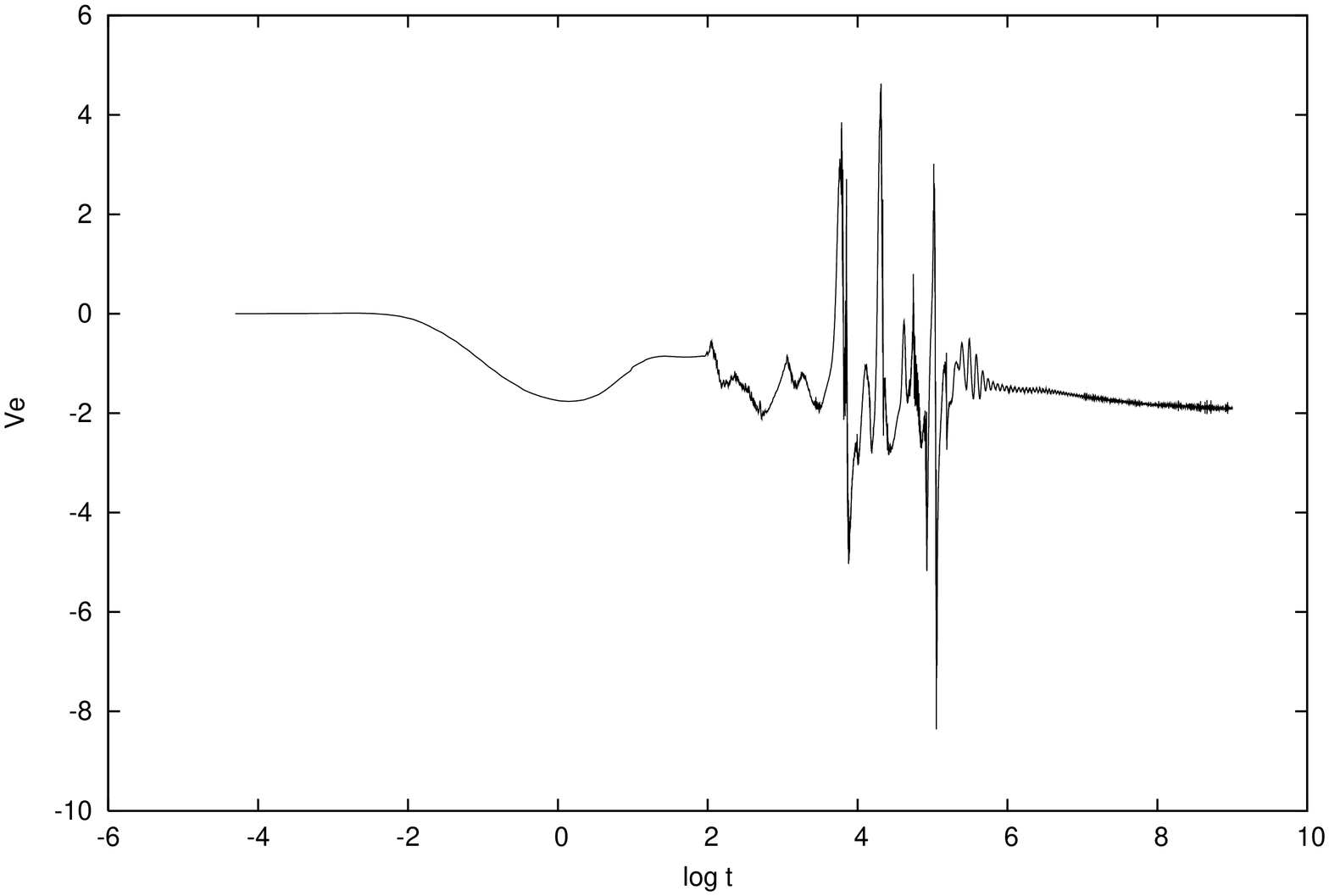, width=6cm, angle=-90}
  \caption{$V_E$ versus time for $\eta=7, \chi=1, \gamma=19,
    \kappa=10$. This state goes finally to the shear state but only
    after a strong clustering instability phase.}
  \label{fig:nuE_eta7_gam19}
\end{figure}

\begin{figure}[h]
  \centering
  \epsfig{file=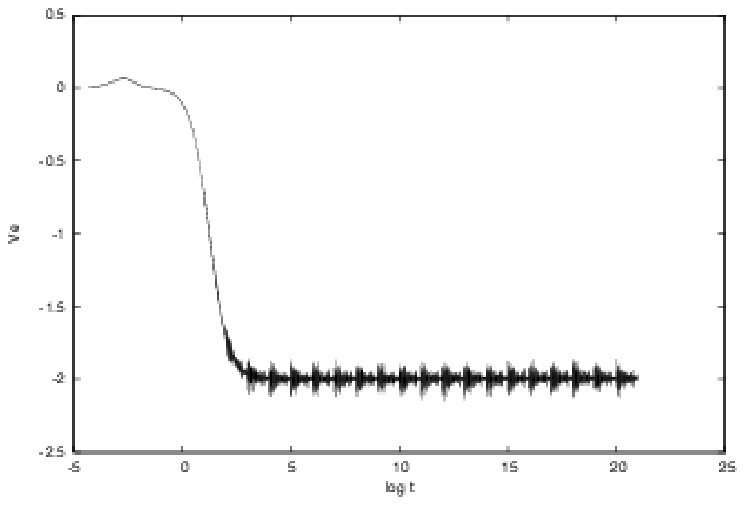, height=6cm}
  \caption{The derivative of the logarithm of the thermal energy with
    respect to the logarithm of the time for a system with $\eta=25,
    \chi=1, \kappa=10, \gamma=0.1$ which does not exit the HCS. We see
    that the derivative settles to -2 and does not change. The small
    deviations from -2 seem to be due to numerical errors.} 
  \label{fig:nuE_eta25_gam01}
\end{figure}

The evolution of the granular thermal energy $E_T = \langle \rho T
\rangle$ has a similar behaviour. The corresponding log-log plots are
shown in figures \ref{fig:hcsett}, \ref{fig:clusterett} and
\ref{fig:vortexett}. It is very interesting to note that the granular
thermal energy $E_T$ does go as $t^{-1}$ for the initial evolution of
the clustering state and this does seem to be independent of the
parameters. This result might be related to the one found by Xiaobo et
al\cite{Xiaobo} and Miller and Luding\cite{Luding2} where the kinetic
energy decays as $t^{-1}$ in molecular dynamics simulations. The slope
of the plots, discussed in some detail in Hill and
Mazenko\cite{ScottHill}, $V_E = \frac{d \log E_T}{d \log t}$, is
another interesting quantity.  Examples of the behaviour of $V_E$ are
shown for several values of the parameters in figures
\ref{fig:nuE_eta25_gam1}, \ref{fig:nuE_eta25_gam19},
\ref{fig:nuE_eta7_gam19} and \ref{fig:nuE_eta25_gam01} respectively.
It is interesting to note that in systems where there is a clustering
instability, $V_E$ has large fluctuations in the later part of the
evolution of this phase.

\begin{figure}[h]
  \centering
  \begin{tabular}{cc}
    \epsfig{file=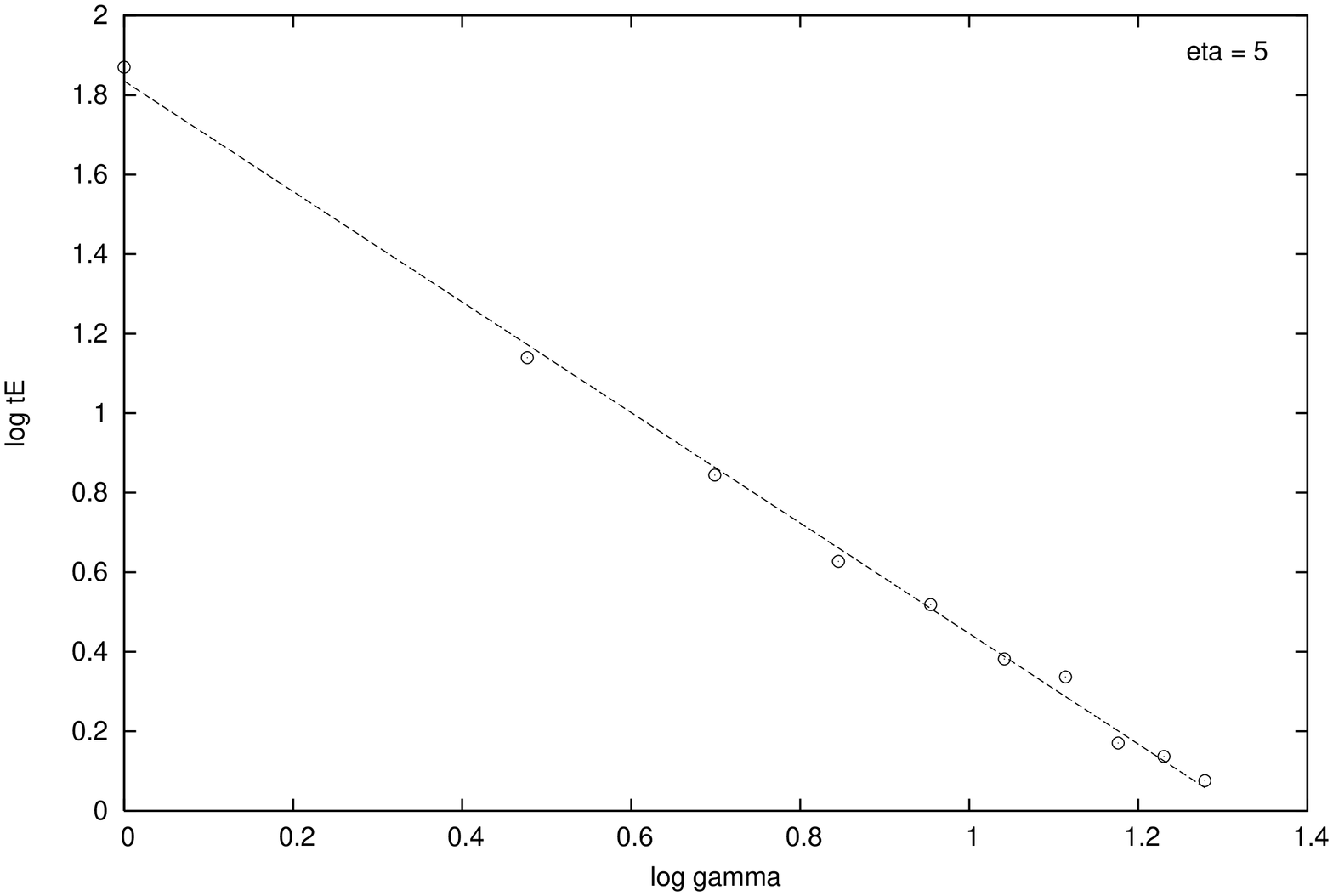, width=5cm, angle=-90} &
    \epsfig{file=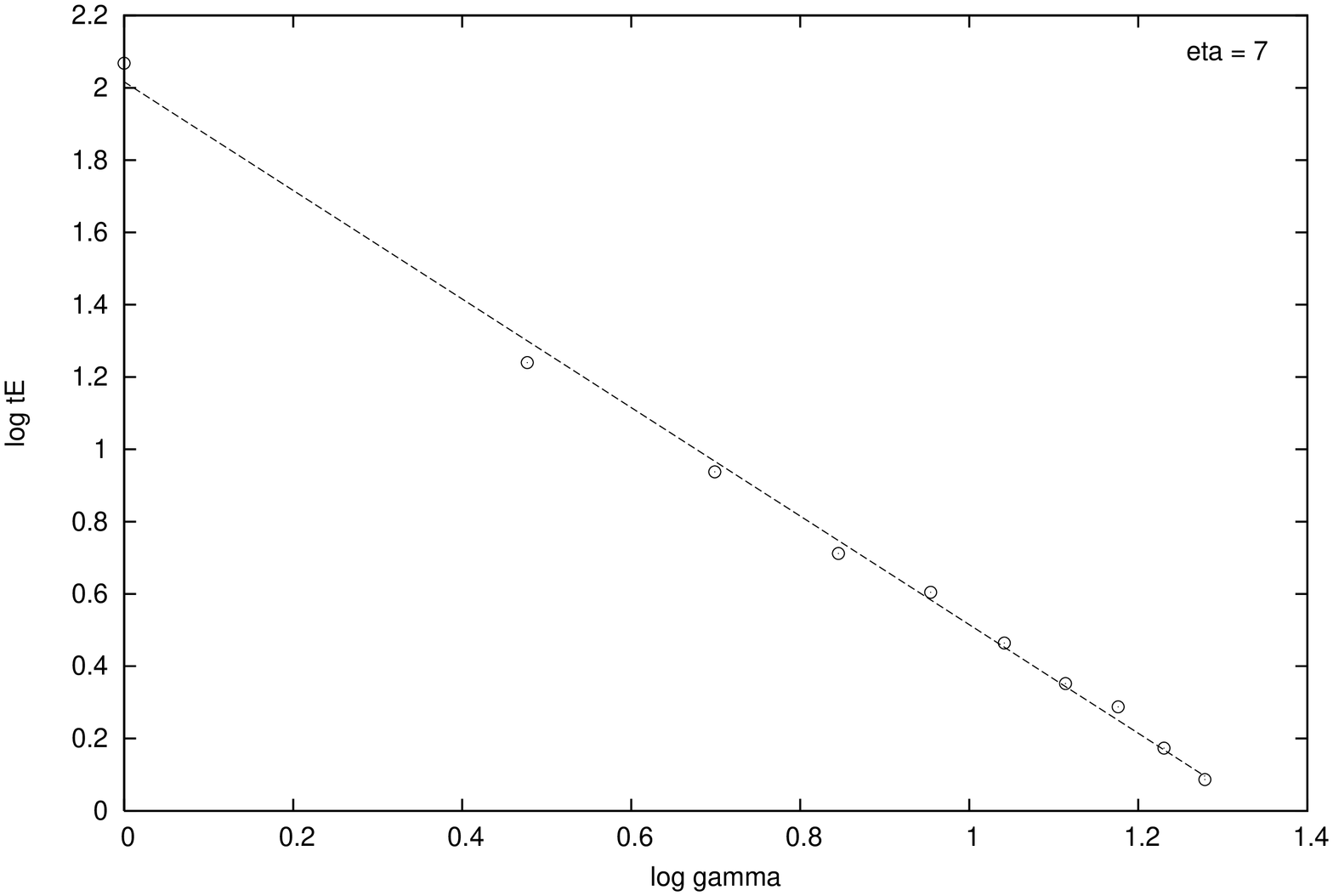, width=5cm, angle=-90}\\
    \epsfig{file=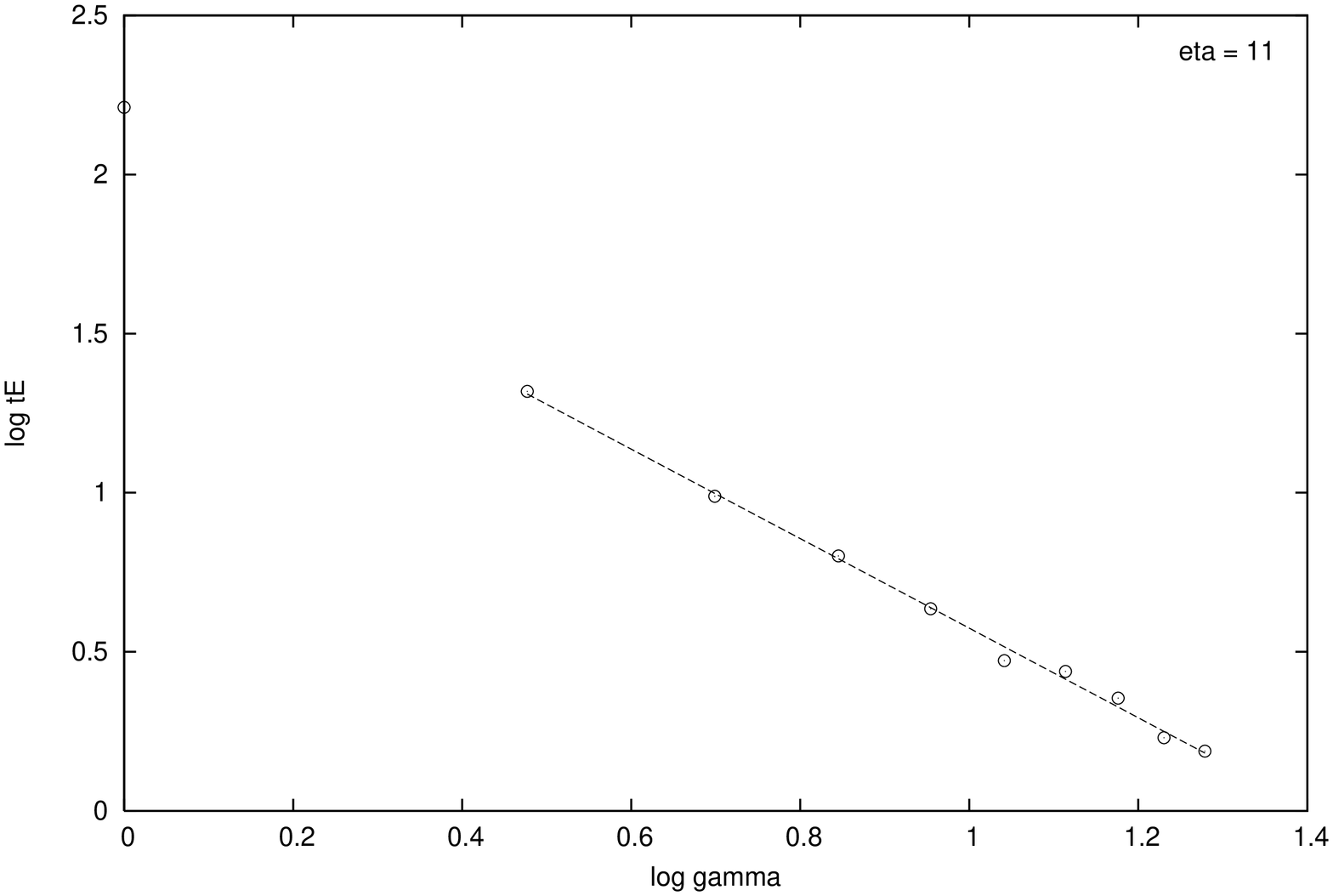, width=5cm, angle=-90} &
    \epsfig{file=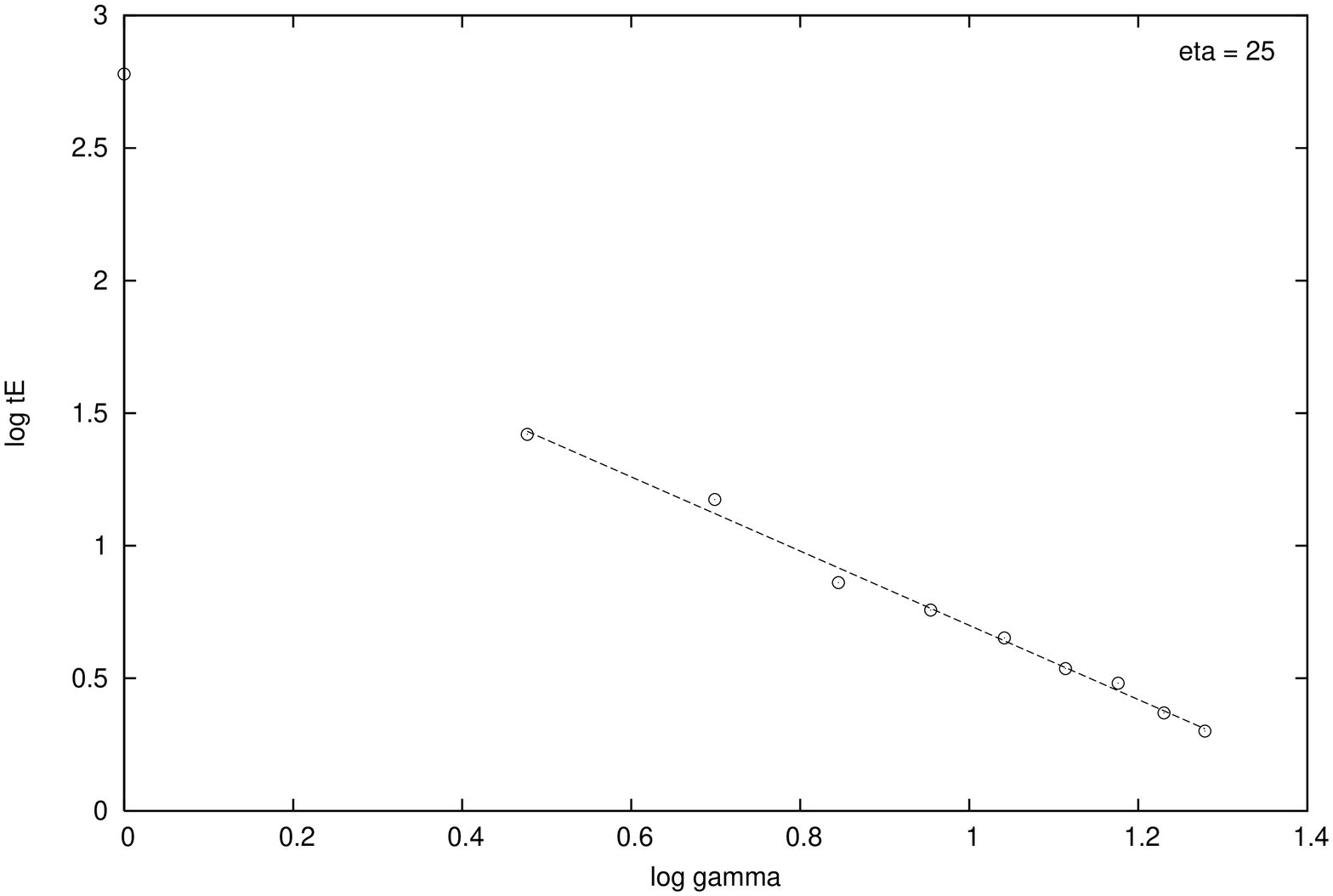, width=5cm, angle=-90} 
  \end{tabular}
  \caption{The power law $t_E = \gamma^\mu$ for $\eta=5, 7, 11,
    25$. Note that for $\eta=11, 25$,  we have to exclude the first
    point from the fit since it is too close to $\gamma_c$. The values
    for $\mu$ are -1.39, -1.50, -1.41 and -1.40 respectively.}
  \label{fig:texitpower}
\end{figure}

Since the energy behaves as (\ref{eq:hcsenergy}) in the HCS and the
system is initially in the HCS, we have 
\begin{equation}
  \label{eq:nue}
  V_E = -\frac{2}{1 + t_0/t}
\end{equation}
Since $t_0$ is relatively small (of order 1), $V_E$ should rapidly
decrease from 0 to -2 if the system stays in the HCS. Any increase in
$V_E$ signals a departure from the HCS and we can hence identify the
time at which $V_E$ reaches a minimum as the time the system begins
to depart from the HCS. We define this time to be $t_{E}$. It was
found earlier\cite{ScottHill} that $t_{E}$\cite{footnote1} followed a
power law with respect to $\gamma$, that is $t_E \sim \gamma^\mu$.
This result cannot hold for all $\gamma$ since, for a given system
size, there is a finite cuttoff $\gamma_c$ below which the system
never leaves the HCS.  However, for $\gamma$ well above this value
(which is generally much less than 1 except for high viscosities), we
find a power law relation for several values of $\eta$ ($\chi=1,
\kappa=10$ being held fixed) with the exponent $\mu$ being in the
range $[-1.5, -1.4]$.  Examples of the power law behavior for several
values of the viscosities are shown in figure \ref{fig:texitpower}. .

\begin{figure}[h]
  \centering
  \epsfig{file=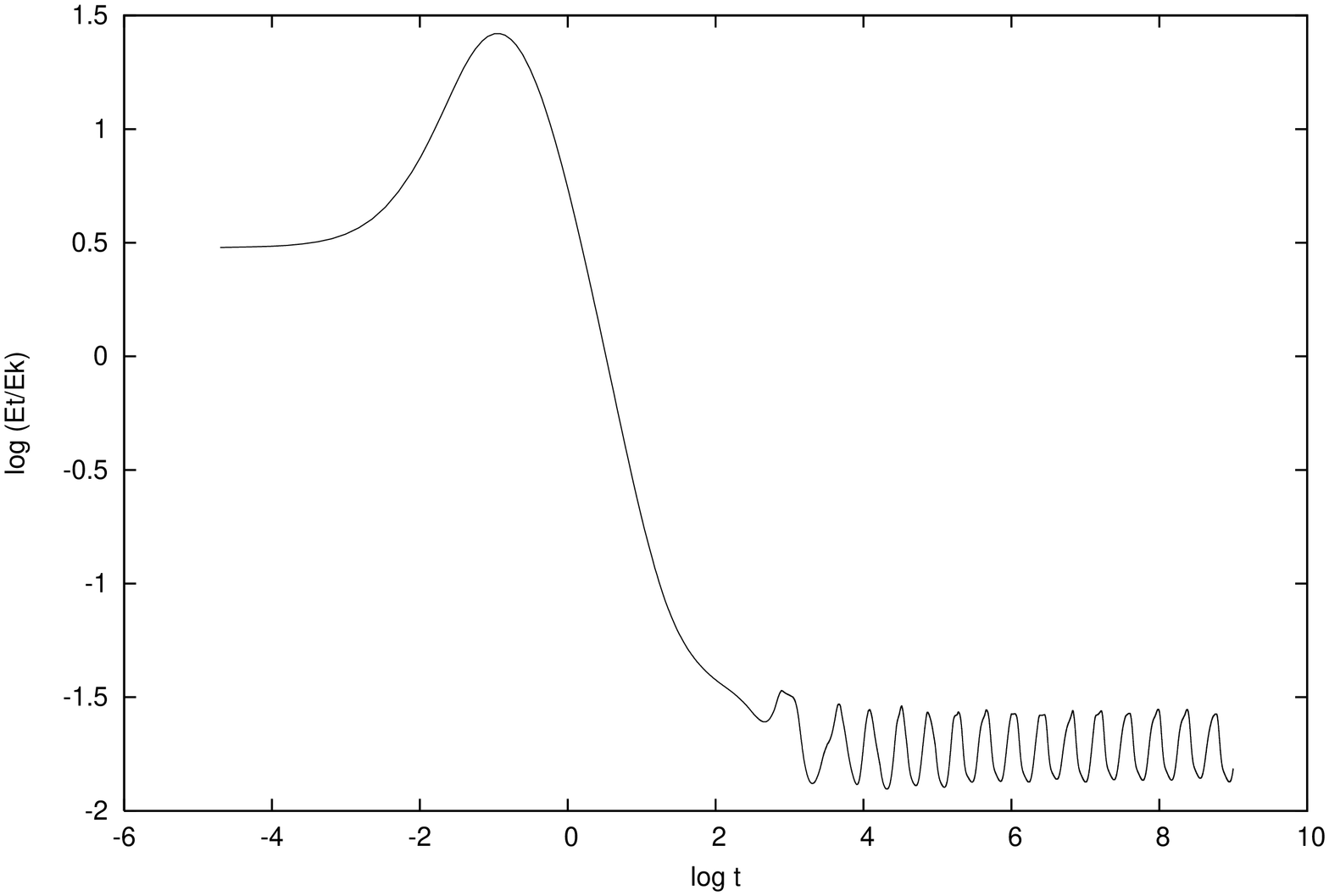, width=6cm, angle=-90}
  \caption{The ratio of the thermal to kinetic energy $E_T/E_k$ for
    $\eta=9, \gamma=11, \chi=1, \kappa=10$ which goes to a final
    vortex state. Note that once the system reaches the vortex state,
    the ratio oscillates about a constant value. } 
  \label{fig:EtEkvortex}
\end{figure}

\begin{figure}[h]
  \centering
  \epsfig{file=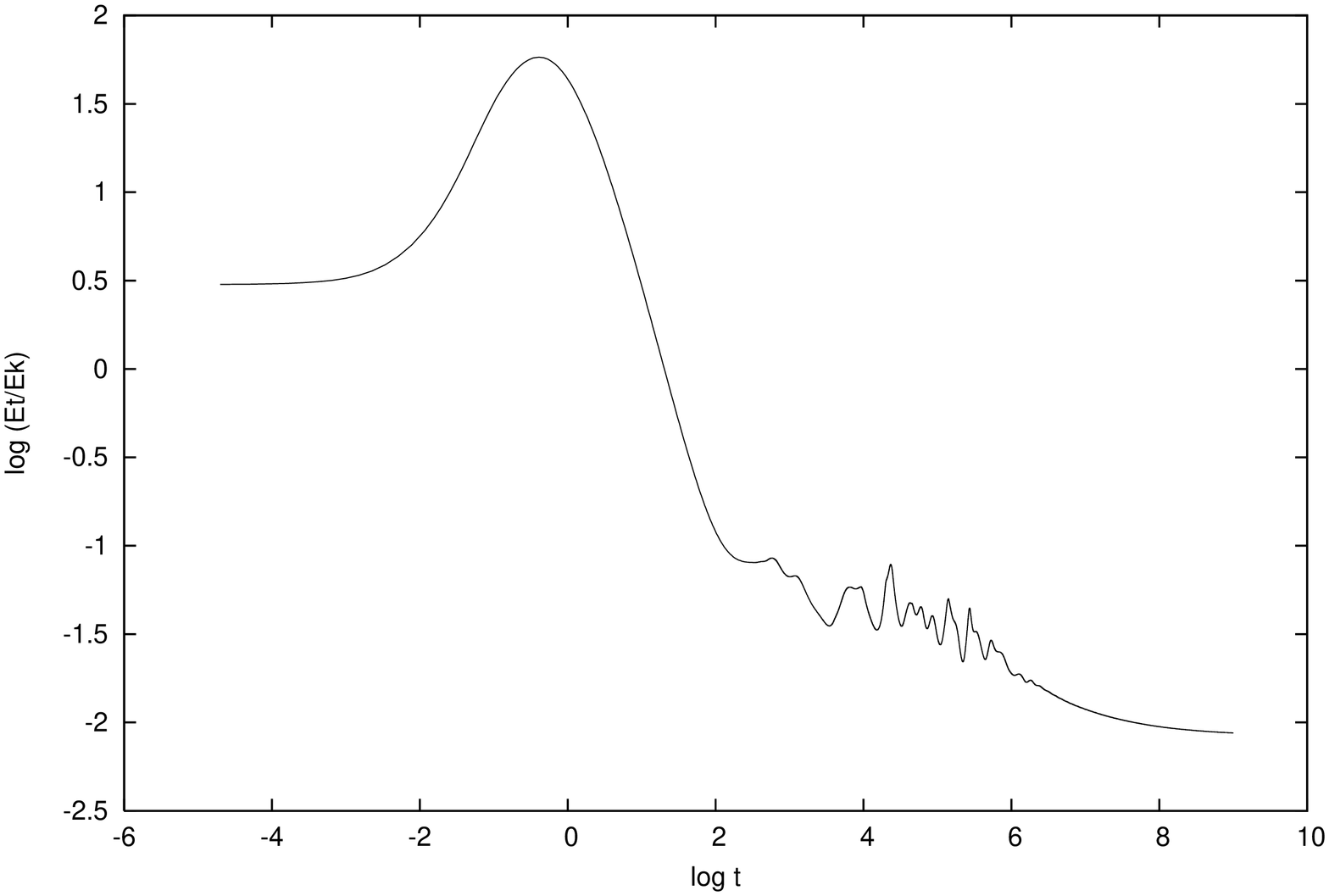, width=6cm, angle=-90}
  \caption{The ratio of the thermal to kinetic energy $E_T/E_k$ for
    $\eta=5, \gamma=3, \chi=1, \kappa=10$ which goes to a final shear
    state. Note that once the system reaches the shear state, the
    ratio converges to a constant. }
  \label{fig:EtEkshear}
\end{figure}

Another quantity of interest which has been analysed earlier in the
literature\cite{McNamara} is the ratio of the thermal energy to the
kinetic energy $E_T/E_k$. For a vortex state, the behaviour is
interesting in that this ratio oscillates about a fixed value as seen
in figure \ref{fig:EtEkvortex}. This is another example of the
cyclical behaviour of this attractor. For a shear state, the ratio
approaches a constant value asymptotically as seen in figure
\ref{fig:EtEkshear}.

\begin{figure}[h]
  \centering
  \epsfig{file=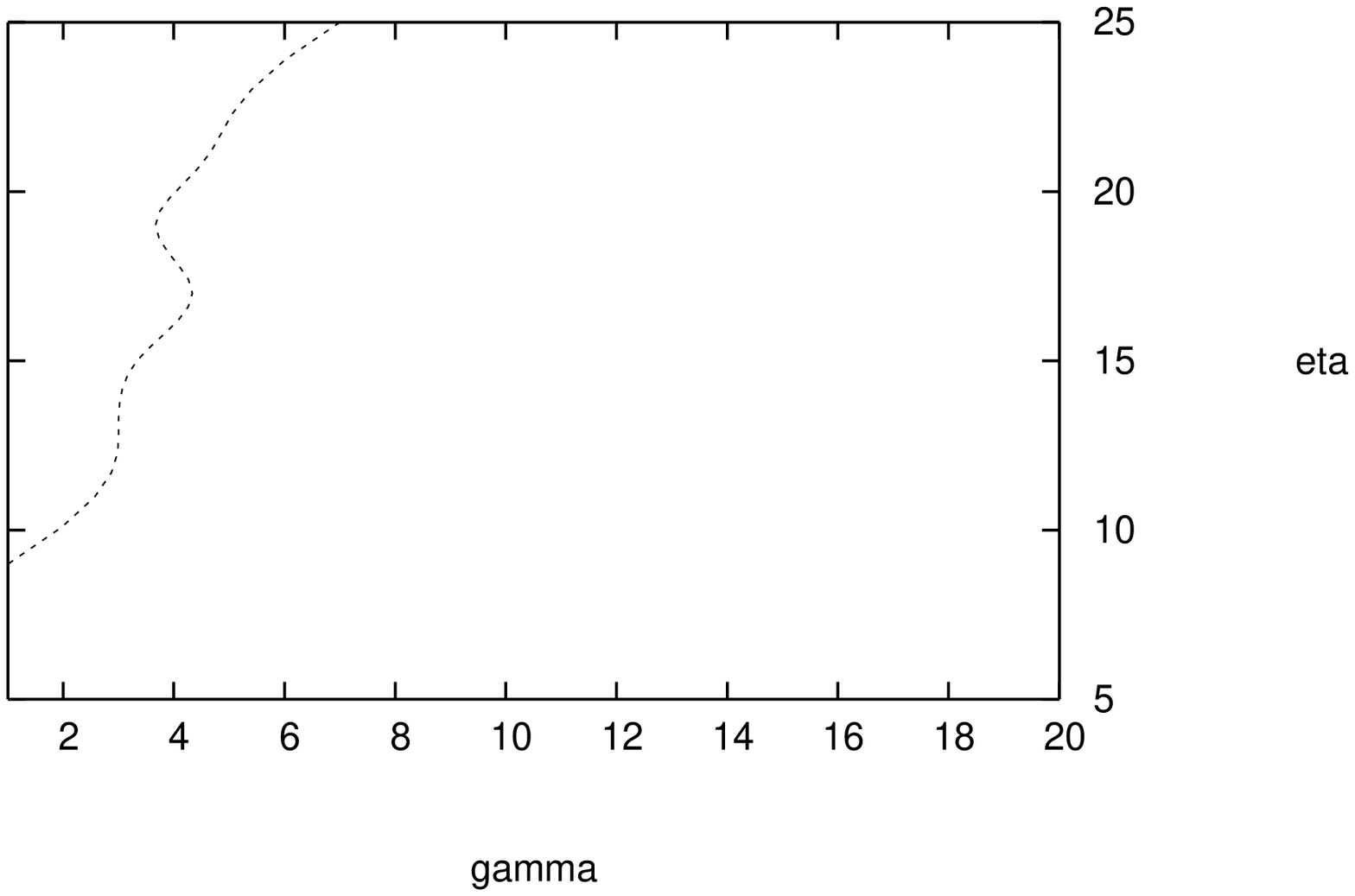, width=6cm, angle=-90}
  \caption{The phase diagram for the clustering instability. The
    values of the other parameters are $\chi=1, \kappa=10$. Clustering
  occurs for parameters to the right of the dashed line.}
  \label{fig:clustering}
\end{figure}

We now look at the clustering instability a bit more closely so as to
identify clearly when it occurs. Since phase separation can be due to
the clustering instability or a result of the velocity gradients in an
attracting state, we have to distinguish between these two. To do so,
we characterize phase separation as having occured if the rms
fluctuation in the density is above 0.3 and the system has reached
the packing density at at least one point. We call the time at which
this first occures as $t_{phase}$. We then identify the time
$t_{final}$ at which the system has reached its final state. The main
feature of the shear band state is the picking of one direction in
which the bands occur. The result of this is a dominance of kinetic
energy in that direction. Hence, for a shear band state, we identify
$t_{final}$ as the first time at which the kinetic energy in one
direction dominates the kinetic energy in the other direction by a
factor of 1000, that is the first time when $\frac{\langle \rho
  u_x^2\rangle}{\langle \rho u_y^2 \rangle} > 1000$ or when
$\frac{\langle \rho u_y^2\rangle}{\langle \rho u_x^2 \rangle} > 1000$.
A similar criterion is used to identify the shearing state in McNamara
and Young\cite{McNamara}. For a vortex state, we indentify $t_{final}$
as the time at which the oscillations in $E_T/E_k$ start.  If
$t_{phase} < t_{final}$ we claim that the phase separation is a result
of the clustering instability.  Using this criterion, the phase
diagram for the clustering instability is shown in figure
\ref{fig:clustering}.

\section{Summary and Conclusions}
To summarize, we have used a hydrodynamic model of granular dynamics
to simulate the cooling of a granular gas with no external field and
with periodic boundary conditions over a large region of parameter
space.  We have found that the results of simulations using this model
are in broad agreement with molecular dynamic simulations of the same
phenomenon.  We find that the system is attracted to one of three
final states - the homogeneous cooling state, the shear band state or
a vortex state.  The HCS is the final state only if the system is
small and the dissipation is very low while the shear band or vortex
states result when the dissipation is high with the exact criterion
for the selection of the state unknown though it has been found to
depend on the initial state of the system.  We have also found a phase
diagram for the clustering instability in this model and have
confirmed the result in Hill and Mazenko\cite{ScottHill} that the time
taken to depart from the homogeneous cooling state follows a power law
with respect to the dissipation coefficient (for high enough
dissipation) over a much larger region of parameter space. It should
be possible to check this result with molecular dynamic simulations.

\section*{Acknowledgements}
We thank Scott Hill for his help and one of us, Srikant Marakani,
would like to thank Mrs. Lalitha Chandrasekhar for providing the
fellowship that enabled him to pursue this research. This work was
also supported by the Materials Science and Engineering Center through
grant No. NSF DMR-9808595. 

\bibliography{outline}

\begin{thebibliography}{10}
\bibitem{footnote3}
The thermal temperature is irrelevant for granular flows since the thermal
  energy scale $k_B T$ is much less than the other energy scales.
\bibitem{footnote5}
In this model, the granular temperature $T$ is an independent variable.
\bibitem{footnote2}
See also Hill and Mazenko\cite{ScottHillone} for a different derivation which
  features the role of a driving free energy.
\bibitem{footnote1}
The quantity used there was $\frac{d \log \langle T \rangle}{d \log t}$ rather
  than $V_E$. This does not produce any significant change in the results. We
  use $V_E$ since $E_T$ is a better behaved quantity compared to the local
  temperature which has large fluctuations with respect to the average when the
  local density is low.
\end{thebibliography}
\bibliographystyle{phaip.bst}
\end{document}